\documentclass[preprintnumbers, floatfix,letterpaper,aps,prd,epsfig,nofootinbib,
twocolumn
]{revtex4-1}
\usepackage{bm,graphicx,dcolumn,epstopdf,epsf, latexsym,mathbbol, amssymb,amsmath,color,slashed, mathrsfs,mathcomp, simplewick}
\usepackage[center]{subfigure}
\usepackage{multirow}
\usepackage{makecell}
\usepackage{tabularray}
\usepackage[colorlinks,linkcolor=blue,citecolor=blue,urlcolor=blue]{hyperref}
\begin{document}
\allowdisplaybreaks
%%%%%%%%%%%%%%%%%%%%%%%%
 \newcommand{\bq}{\begin{equation}}
 \newcommand{\eq}{\end{equation}}
 \newcommand{\bqn}{\begin{eqnarray}}
 \newcommand{\eqn}{\end{eqnarray}}
 \newcommand{\nb}{\nonumber}
 \newcommand{\lb}{\label}
 \newcommand{\f}{\frac}
 \newcommand{\p}{\partial}
%%%%%%%%%%%%%%%%%%%%%%%%%
\newcommand{\PRL}{Phys. Rev. Lett.}
\newcommand{\PLB}{Phys. Lett. B}
\newcommand{\PRD}{Phys. Rev. D}
\newcommand{\CQG}{Class. Quantum Grav.}
\newcommand{\JCAP}{J. Cosmol. Astropart. Phys.}
\newcommand{\JHEP}{J. High. Energy. Phys.}
\newcommand{\red}{\textcolor{black}}
 %%%%%%%%%%%%%%%%%%%%%%%%
%

\title{Interpretable Machine Learning Strategies for Accurate Prediction of Thermal Conductivity in Polymeric Systems}

\author{Chunbo Lin${}^{a}$}
\email{corresponding author: 202103170412@zjut.edu.cn}

\author{Han Zheng${}^{b}$}

\affiliation{
${}^{a}$ College of Science, Zhejiang University of Technology, Hangzhou, 310032, China\\
${}^{b}$ College of Textile Science and Engineering (International Institute of Silk), Zhejiang Sci-Tech University, Hangzhou, 310018, China \\
}

%\date{\today}
\date{March 5, 2024}

\begin{abstract}
	Polymers, integral to advancements in high-tech fields, necessitate the study of their thermal conductivity (TC) to enhance material attributes and energy efficiency. The TC of polymers obtained by molecular dynamics (MD) calculations and experimental measurements is slow, and it is difficult to screen polymers with specific TC in a wide range. Existing machine learning (ML) techniques for determining polymer TC suffer from the problems of too large feature space and cannot guarantee very high accuracy. In this work, we leverage TCs from accessible datasets to decode the Simplified Molecular Input Line Entry System (SMILES) of polymers into ten features of distinct physical significance. A novel evaluation model for polymer TC is formulated, employing four ML strategies. The Gradient Boosting Decision Tree (GBDT)-based model, a focal point of our design, achieved a prediction accuracy of R$^2$ = 0.88 on a dataset containing 400 polymers. Furthermore, we used an interpretable ML approach to discover the significant contribution of quantitative estimate of drug-likeness and number of rotatable bonds features to TC, and analyzed the physical mechanisms involved. The ML method we developed provides a new idea for physical modeling of polymers, which is expected to be generalized and applied widely in constructing polymers with specific TCs and predicting all other properties of polymers.
\end{abstract}

%\pacs{98.80.Cq, 98.80.Qc, 04.50.Kd, 04.60.Bc}

\maketitle
\section{Introduction}
\renewcommand{\theequation}{1.\arabic{equation}} \setcounter{equation}{0}

Polymers play a crucial role in today's world, finding uses in advanced areas like implantable brain-computer interfaces, electronic chips, and wearable technologies.\cite{1} Polymers characterized by elevated thermal conductivity (TC) are instrumental in augmenting the heat dissipation capacity of devices, thereby mitigating the potential adverse impacts of overheating on device functionality or user comfort\cite{2}. Conversely, polymers exhibiting reduced thermal conductivity harness exceptional thermal insulation attributes, finding extensive utilization in thermal insulation applications, such as within construction sector walls and thermal management systems for electronic devices, aiming to diminish heat loss and enhance energy efficiency. The identification of polymers with specific thermal conductivities represents a noteworthy endeavor.
\setlength{\parskip}{4pt}

However, the current dominant approaches for screening polymers with specific thermal conductivities are molecular dynamics (MD) calculations\cite{3} or experimental measurements\cite{1}. Polymeric systems, characterized by their extensive size and substantial atomic count, render the application of MD methodologies for the calculation of TC inefficient. Experimentally ascertaining the TC of polymers necessitates intricate procedures, such as meticulous sample preparation and rigorous regulation of environmental variables, thereby imposing stringent demands on the precision of experimental methodologies.\cite{5} Such methodologies demand substantial temporal investments, extending from days to weeks, to deduce the TC features of complex polymers, often yielding results with error margins that may be deemed unsatisfactory. To efficiently sift through a diverse array of materials for particular thermal conductivities, an urgent requirement emerges for a rapid and precise technique to forecast the TC of polymers.

\setlength{\parskip}{4pt}
In recent years, machine learning (ML) has witnessed a significant surge in its application, demonstrating remarkable success in achieving high levels of accuracy in forecasting outcomes such as carbon dioxide emissions\cite{6} and properties of organic solar cells\cite{7}. Previous studies\cite{8} have leveraged a ML paradigm to estimate the TC of materials. They compiled a dataset comprising 469 amorphous polymers, converting the polymer-Simplified Molecular Input Line Entry System (p-SMILES) into 300-dimensional continuous value vectors, which yielded a prediction accuracy with a coefficient of determination (R$^2$) of 0.828. However, the complexity of the input features and the employment of 300-dimensional vectors, derived through linguistic processing devoid of physical significance, resulted in predictive performance that did not meet expectations.

\setlength{\parskip}{4pt}
We decode the p-SMILES of polymers into 10 features imbued with physical significance, thereby shrinking the feature space by a factor of 30 relative to the previous methodology\cite{7}. Our approach entails the construction of a ML model predicated on Gradient Boosting Decision Trees (GBDT) for the estimation of polymers' TC, culminating in an enhanced model accuracy with a coefficient of determination (R$^2$) of 0.93. 

\setlength{\parskip}{4pt}
Furthermore, the features integrated into our model are intrinsically interpretable, laying the groundwork for interpretable analyses of the predictive framework. We have elucidated a series of characteristic polymer attributes, such as the number of rotatable bonds and the quantitative estimate of drug-likeness, that significantly influence the thermal conductivity of polymers. Furthermore, we have delineated the physical mechanisms underpinning the associations between these attributes and thermal conductivity. The ML methodology delineated in this study introduces a conceptual framework for the modeling of polymers, capable of predicting not merely the TC but encompassing all pertinent properties of the polymer under investigation.

\section{METHODS}
\renewcommand{\theequation}{2.\arabic{equation}} \setcounter{equation}{0}

\begin{figure}[htbp]
	\centering
	\includegraphics[width=0.5\textwidth]{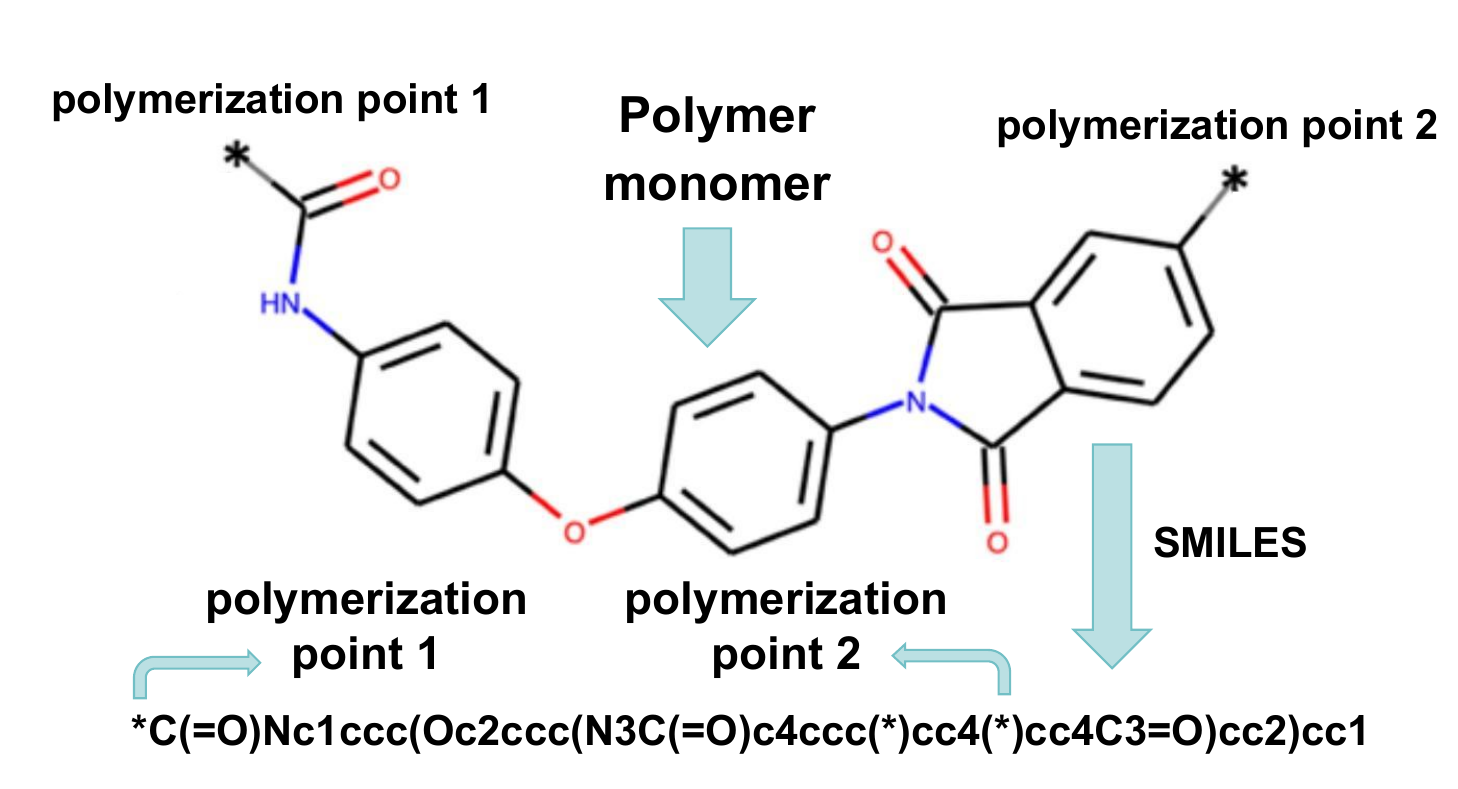}
	\caption{Illustration of a SMILES notation expressed as a string for depicting the molecular architecture of a polymer.}
	\label{Fig1}
\end{figure}

We employed the third-party libraries \textit{RadonPy}\cite{9} and \textit{RDKit}\cite{11}, sourced from GitHub, as our datasets. RadonPy comprises the Simplified Molecular Input Line Entry System (SMILES)\cite{12} representations for 1077 polymers alongside TC data computed through MD methodologies. The SMILES notation for polymers typically encapsulates recurring monomeric units, contingent upon the structure and composition of monomers within the polymer. An exemplar is illustrated in Figure \ref{Fig1}. 

%\begin{figure}[htbp]
%	\centering
%%	\includegraphics[scalebox={-1}{1}, width=0.5\textwidth]{figure1R.pdf} % 水平镜像并指定宽度
%%	\reflectbox{\includegraphics[width=0.5\textwidth]{figure1R.pdf}}
%    \scalebox{1}[-1]{\includegraphics[height=4.5cm]{figure1R.pdf}}
%	\caption{Illustration of a SMILES notation expressed as a string for depicting the molecular architecture of a polymer.}
%	\label{Fig1}
%\end{figure}

\setlength{\parskip}{4pt}
In this study, we opted to train our model using the initial 400 polymers from the dataset. The MolecularDescriptors module within the \textit{RDKit} library facilitated the extraction of characteristic parameters imbued with physical significance for each polymer. These parameters were derived from the decoding of polymer SMILES into 10 eigenvectors (eg., Number of Rotatable Bonds\cite{13}), thereby constituting a 10-dimensional feature space. The determination of TC in polymers conventionally necessitates weeks or even months of experimental measurements or alternatively, days of MD simulation calculations\cite{14}. Consequently, to streamline the screening process for polymers possessing targeted TC, we opted to devise a regression model to delineate the relationship between SMILES representations and TC. This approach enables the accurate prediction of polymer TC within a significantly shorter timeframe.

\setlength{\parskip}{4pt}
We utilized \textit{scikit-learn}\cite{15}, a \textit{Python} library renowned for its capabilities in ML, to conduct training on Multi-layer Perceptron (MLP)\cite{16}, Random Forest (RF)\cite{17}, and GBDT\cite{18} models. Additionally, for training eXtreme Gradient Boosting (XGBoost)\cite{19} model, we employed the \textit{Python} library \textit{xgboost}\cite{20}. Following rigorous experimentation, we identified the GBDT model as yielding the most favorable training results among the four models examined. Hyperparameter tuning facilitated the optimization of key parameters, with the Number of trees set to 300, Maximum depth of each tree set to 5, Minimum number of samples required to split a node set to 4, Minimum number of samples required at each leaf node set to 1, Learning rate set to 0.01, and Subsample ratio set to 0.9.

\setlength{\parskip}{4pt}
We operate under the assumption that the dataset obtained is accurate and that the TC computed through MD simulation represents the authentic TC of the polymer. Moreover, given that predictions of polymer TC are based solely on monomer information, it is presupposed that elements such as the degree of polymerization, temperature, and the spatial configuration of the polymer are considered ancillary influences on polymer TC.

\section{RESULTS AND DISCUSSION}
\renewcommand{\theequation}{3.\arabic{equation}} \setcounter{equation}{0}

In the present investigation, we transmute the SMILES notation into a ten-dimensional attribute sphere. The nomenclature for each feature within this multidimensional expanse, as delineated by the \textit{RDKit} computational library, is cataloged in the inaugural column of Table \ref{Tab1}. Additionally, Table \ref{Tab1} elucidates the physical significances and metrications of these features. For succinctness, the abbreviations denoting the physical properties of these polymers, as presented in the secondary column of Table \ref{Tab1}, will henceforth represent these features. The selected features encompass: the molecular weight's mean value (MWT), the Quantitative Estimate of Drug-likeness (QED), the molecule's valence electron count (NVE), the computation of Balaban's J metric (BBJ), the molecule's total surface area (TPS), the tally of Hydrogen Bond Acceptors (NHA), the count of Rotatable Bonds (NRB), the Wildman-Crippen LogP valuation (MLP), the Wildman-Crippen MR valuation (MMR), and the enumeration of halogen elements (FHA).

\begin{table*}[htbp]
	\centering
	\caption{Ten features from SMILES for training applications}
	\label{Tab1}
	\begin{tblr}{
			cells = {c},
			hline{1-2,12} = {-}{},
		}
		\textbf{\textbf{Feature (in }\textit{\textbf{\textit{RDKit}}}\textbf{)}} & \textbf{\textbf{Abbreviation}} & \textbf{\textbf{Physical meaning}}               & \textbf{\textbf{Unit}} \\
		MolWt                                                                    & MWT                            & The average molecular weight of the molecule     & amu                    \\
		qed                                                                      & QED                            & Quantitative Estimate of Drug-likeness           & \textbackslash{}       \\
		NumValenceElectrons                                                      & NVE                            & The number of valence electrons the molecule has & \textbackslash{}       \\
		BalabanJ                                                                 & BBJ                            & Calculate Balaban's J value for a molecule       & \textbackslash{}       \\
		TPSA                                                                     & TPS                            & The total surface area of a molecule             & Å²                     \\
		NumHAcceptors                                                            & NHA                            & Number of Hydrogen Bond Acceptors                & \textbackslash{}       \\
		NumRotatableBonds                                                        & NRB                            & Number of Rotatable Bonds                        & \textbackslash{}       \\
		MolLogP                                                                  & MLP                            & Wildman-Crippen LogP value                       & \textbackslash{}       \\
		MolMR                                                                    & MMR                            & Wildman-Crippen MR value                         & cm³/mol                \\
		fr\_halogen                                                              & FHA                            & Number of halogens                               & \textbackslash{}       
	\end{tblr}
\end{table*}

\begin{figure*}
	\centering
	\begin{minipage}{0.5\textwidth}
		\centering
		\includegraphics[width=\textwidth]{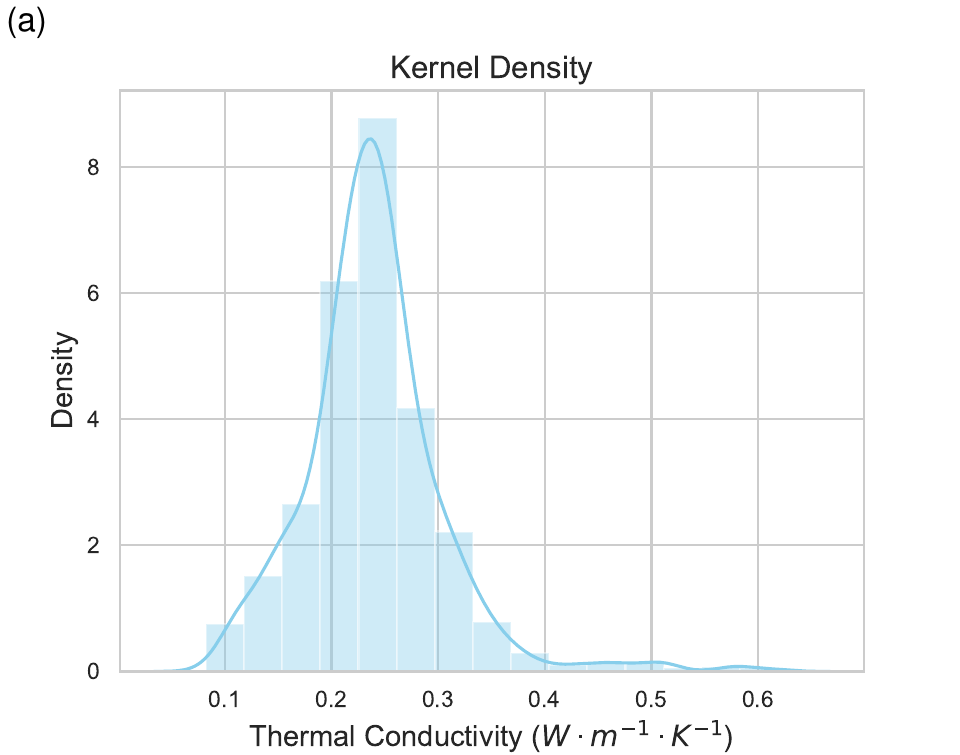}
		%		\caption{a}
		\label{Fig2a}
	\end{minipage}%
	\begin{minipage}{0.5\textwidth}
		\centering
		\includegraphics[width=\textwidth]{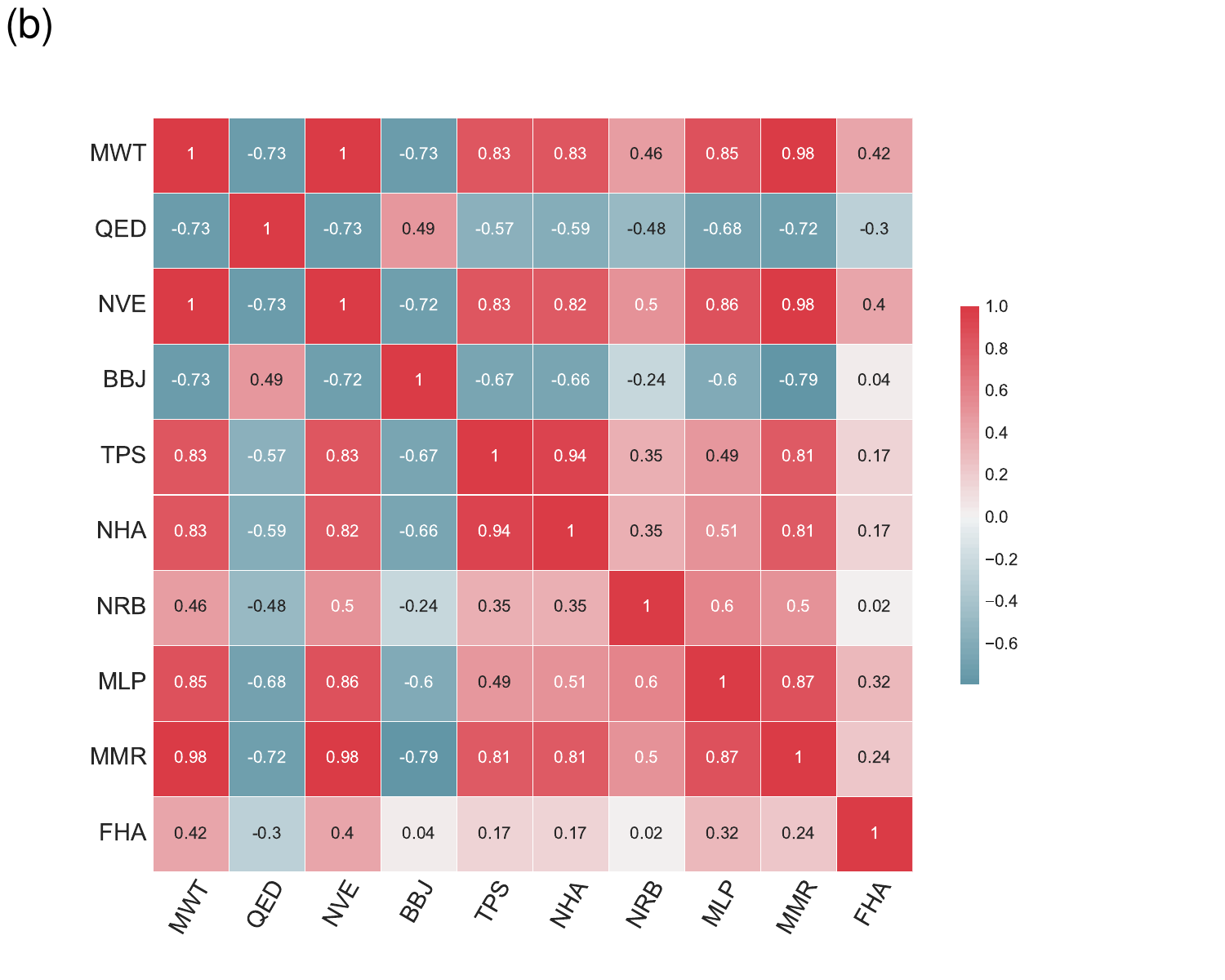}
		%		\caption{b}
		\label{Fig2b}
	\end{minipage}
	\caption{(a) Kernel density distribution of polymer TC within the dataset; (b) Heatmap depicting pearson correlation coefficients among ten distinct features.}
	\label{Fig2}
\end{figure*}

\setlength{\parskip}{4pt}
To evaluate the comprehensive distribution of TC among the polymers encompassed in our dataset, Figure \ref{Fig2}a was constructed. This figure, employing a kernel density estimation technique, delineates TC on the x-axis, with values spanning approximately from 0.06 to 0.7 W$\cdot$m\textsuperscript{-1}$\cdot$K\textsuperscript{-1}, while the y-axis quantifies the density of the numerical simulation. The prominence of the blue curve at any locus within Figure \ref{Fig2}a signifies the aggregation of data points proximal to that specific value of TC. Manifesting a broadly symmetrical bell-shaped curve, and with the dataset affirming conformity to a normal distribution as evidenced by the Shapiro-Wilk test, it is posited that our training specimens are normally distributed.

\begin{figure*}[htbp]
	\centering
	\includegraphics[width=0.8\textwidth]{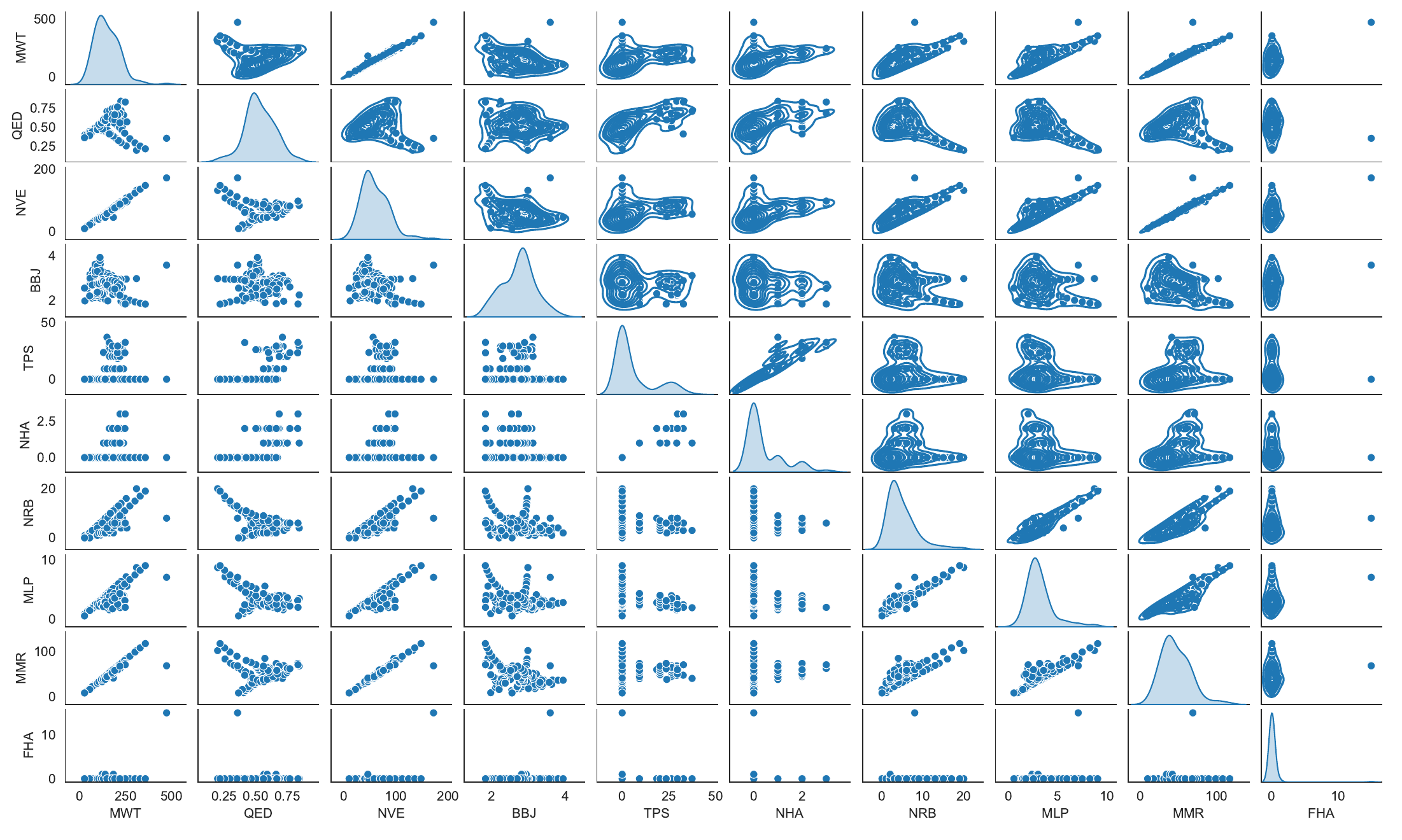}
	\caption{Pairwise correlation matrix with kernel density estimations for selected features.}
	\label{Fig3}
\end{figure*}

\setlength{\parskip}{4pt}
To mitigate the inclusion of superfluous data ensuing from highly correlated features and to preclude inefficiencies during model training, Figure \ref{Fig2}b was devised. This figure illustrates the Pearson correlation coefficients encapsulated within each square, quantifying the interrelation between pairs of features. The intensity of each square's hue signifies the correlation level between the corresponding features, with darker shades indicating higher correlation (dark red for positive, dark blue for negative) and lighter shades denoting weaker correlation. A pronounced correlation is notably observed between MWT and MMR, as well as NVE and MMR, each registering a coefficient of 0.98. Despite this, a deeper examination reveals that MMR, MWT, and NVE encapsulate distinct physical properties. To ensure no pertinent information is overlooked, we opted to retain both MWT and NVE within our feature set. Our analysis of the ten-dimensional feature space reveals a scarcity of highly correlated features, with the majority displaying negligible correlation. This indicates the selected features possess intrinsic value, underscoring the dataset's overall rationality.

\setlength{\parskip}{4pt}
We employed pairwise relationship plots to elucidate the bivariate relationships among features, as well as the distribution of individual features within the multivariate dataset, as depicted in Figure \ref{Fig3}. For illustrative purposes, only the first 150 data points from the dataset were selected for plotting. The ten plots residing on the diagonal of this scatterplot matrix represent kernel density estimation plots for single features, delineating the distribution of each feature in isolation. Each blue point within Figure \ref{Fig3} symbolizes a polymer sample. Off-diagonal grids showcase small plots that elucidate the relationship between features labeled on the rows and those on the columns. For instance, the second plot in the first row elucidates the relationship between the MWT and QED features. This scatterplot matrix features a scatterplot in its lower left quadrant and a contour plot in the upper right, with contours illustrating the data's sparsity. The dataset contains a minimal number of anomalous samples, which were substituted with other normal samples from the dataset. At this juncture, the exploratory data analysis phase preceding ML modeling has been concluded.

\begin{table*}[htbp]
	\centering
	\caption{Hyperparameter optimization results for different models}
	\label{Tab2}
	\begin{tblr}{
			cells = {c},
			vline{3,5,7} = {-}{},
			hline{1-2,9} = {-}{},
		}
		\textbf{\textbf{MLP}}                  & \textbf{\textbf{Value}} & \textbf{\textbf{GBDT}}                & \textbf{\textbf{Value}} & \textbf{\textbf{XGBoost}}            & \textbf{\textbf{Value}} & \textbf{\textbf{RF}}                  & \textbf{\textbf{Value}} \\
		\textit{\textit{hidden\_layer\_sizes}} & (50, 50)                & \textit{\textit{n\_estimators}}       & 300                     & \textit{\textit{n\_estimators}}      & 100                     & \textit{\textit{n\_estimators}}       & 150                     \\
		\textit{\textit{alpha}}                & 0.005                   & \textit{\textit{max\_depth}}          & 5                       & \textit{\textit{max\_depth}}         & 5                       & \textit{\textit{max\_depth}}          & 6                       \\
		\textit{\textit{tol}}                  & 0.0001                  & \textit{\textit{min\_samples\_split}} & 4                       & \textit{\textit{min\_child\_weight}} & 1                       & \textit{\textit{min\_samples\_split}} & 2                       \\
		\textit{\textit{max\_iter}}            & 300                     & \textit{\textit{min\_samples\_leaf}}  & 1                       & \textit{\textit{colsample\_bytree}}  & 0.7                     & \textit{\textit{min\_samples\_leaf}}  & 1                       \\
		\textit{\textit{learning\_rate\_init}} & 0.01                    & \textit{\textit{learning\_rate}}      & 0.01                    & \textit{\textit{learning\_rate}}     & 0.05                    & \textbackslash{}                      & \textbackslash{}        \\
		\textit{\textit{momentum}}             & 0.9                     & \textit{\textit{subsample}}           & 0.9                     & \textit{\textit{subsample}}          & 0.7                     & \textbackslash{}                      & \textbackslash{}        \\
		\textit{\textit{validation\_fraction}} & 0.1                     & \textbackslash{}                      & \textbackslash{}        & \textbackslash{}                     & \textbackslash{}        & \textbackslash{}                      & \textbackslash{}        
	\end{tblr}
\end{table*}

\setlength{\parskip}{4pt}
We first normalized the dataset, which comprises feature spaces and TC of 400 polymers, yielding a novel dataset encompassing 4400 data points. Subsequently, this dataset was subjected to training employing four distinct ML models: MLP, RF, GBDT, and XGBoost. As delineated in the METHODS section, the training outcomes of these models are significantly influenced by their hyperparameters, which consequently affect the accuracy of TC predictions. We engaged in the selection of six hyperparameters for the GBDT model, subjecting it to training across a spectrum of hyperparameter values. These hyperparameters include the number of weak learners (\texttt{n\_estimators}), the maximum depth of the tree (\texttt{max\_depth}), the minimum number of samples required to split a node (\texttt{min\_samples\_split}), the minimum number of samples required at a leaf node (\texttt{min\_samples\_leaf}), the learning rate (\texttt{learning\_rate}), and the subsample ratio of the training instance (\texttt{subsample}).

\begin{figure*}[htbp]
	\centering
	\begin{minipage}[b]{0.5\textwidth}
		\centering
		\includegraphics[width=0.8\textwidth]{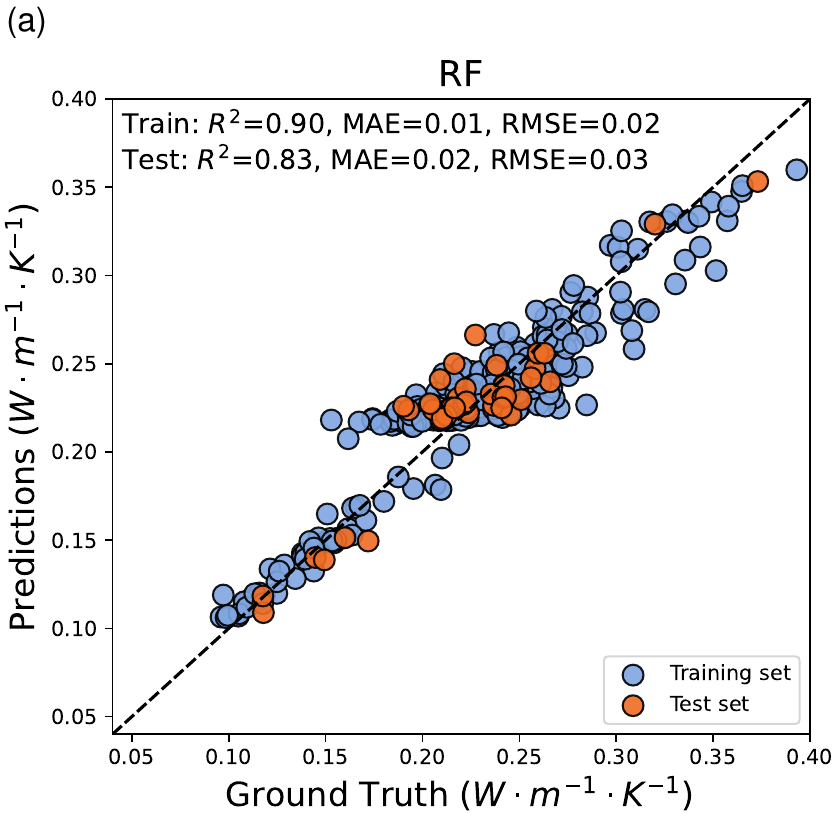}
		%\caption*{(a) Kernel density distribution of polymer TC within the dataset.}
		\label{Fig4a}
	\end{minipage}%
	\begin{minipage}[b]{0.5\textwidth}
		\centering
		\includegraphics[width=0.8\textwidth]{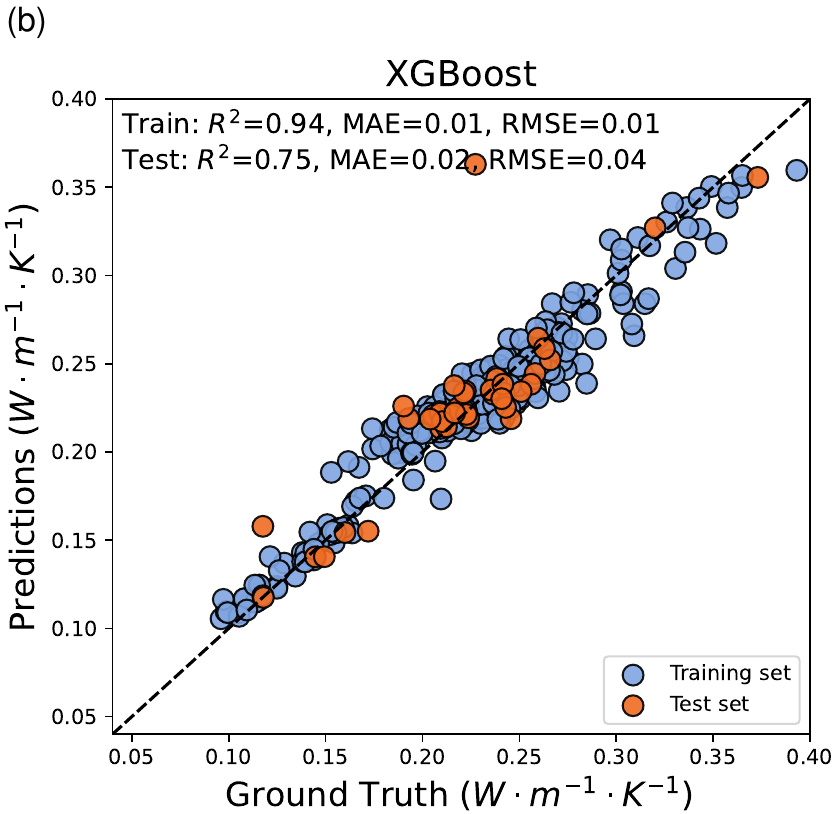}
		%\caption*{(b) Heatmap depicting Pearson correlation coefficients among ten distinct features.}
		\label{Fig4b}
	\end{minipage}
	
	\vspace{0cm} % 适当调整两行图片的垂直间距
	
	\begin{minipage}[b]{0.5\textwidth}
		\centering
		\includegraphics[width=0.8\textwidth]{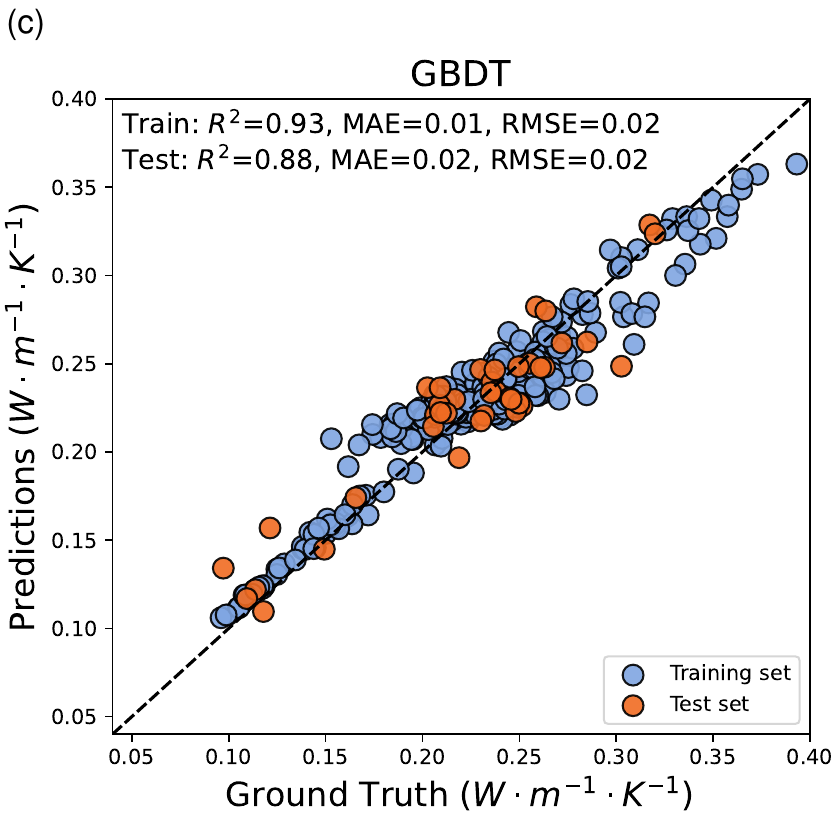}
		%		\caption*{(c) Another figure caption.}
		\label{Fig4c}
	\end{minipage}%
	\begin{minipage}[b]{0.5\textwidth}
		\centering
		\includegraphics[width=0.8\textwidth]{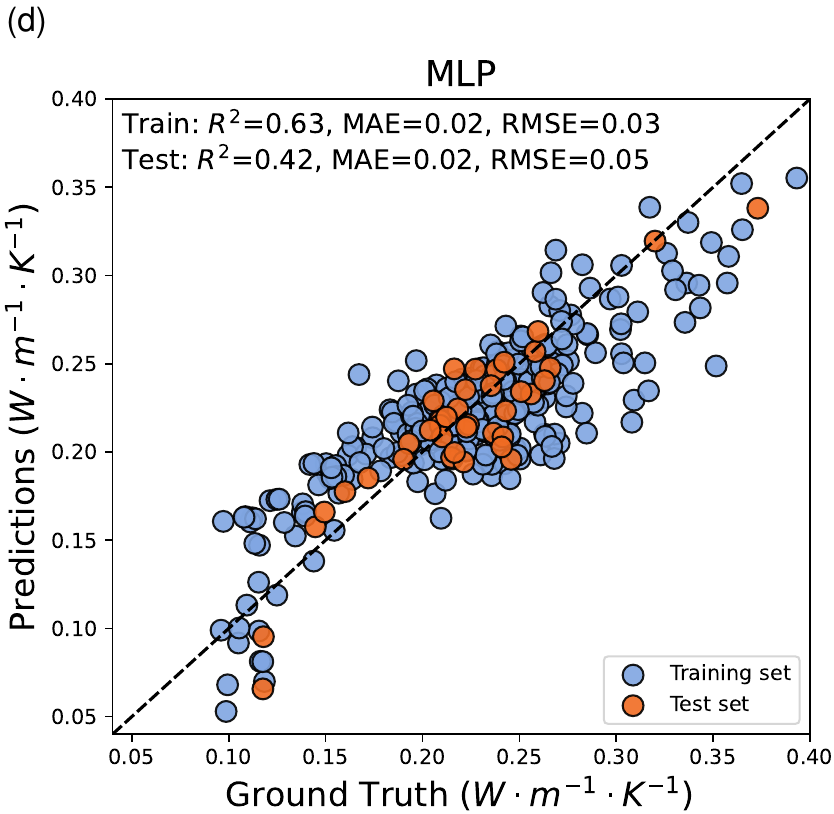}
		%		\caption*{(d) Yet another figure caption.}
		\label{Fig4d}
	\end{minipage}
	
	\caption{Comparative pairwise plots of predicted versus ground truth TC, as calculated by MD, across training and test datasets for four models: MLP, RF, GBDT, and XGBoost, with evaluation metrics including R2, Mean Absolute Error (MAE), and Root Mean Square Error (RMSE).}
	\label{Fig4}
\end{figure*}

\setlength{\parskip}{4pt}
To mitigate variability in training outcomes attributable to differing data partitioning approaches, thereby enhancing the stability and reliability of our estimations, we implemented 10-fold cross-validation during the hyperparameter optimization process. Preliminary training sessions revealed a propensity for overfitting within the model. To augment the model's generalization capability, introduce greater stochasticity, diminish its sensitivity to noise in the training data, and thus counteract overfitting, we judiciously decreased the number of trees and the maximum tree depth while increasing the minimum number of samples required for both node splitting and leaf nodes. Throughout the training phase, a grid search was employed to meticulously explore the hyperparameter space for all six hyperparameters, with the objective of identifying the most efficacious hyperparameter combination. For the MLP model, seven hyperparameters were optimized, whereas the RF model's optimization involved four hyperparameters, and the XGBoost model was optimized across six hyperparameters. The outcomes of hyperparameter optimization for these models are systematically cataloged in Table \ref{Tab2}.

\setlength{\parskip}{4pt}
Figure \ref{Fig4} delineates the juxtaposition of the Predicted versus Ground Truth values across four ML models post hyperparameter optimization. The abscissa represents the true TC values of polymers, ascertained via MD simulations, while the ordinate corresponds to the models' TC predictions. The dataset was partitioned into a test subset, depicted by orange dots (10\%), and a training subset, illustrated by purple dots (90\%). Each subplot's header enumerates the metrics R², MAE, and RMSE, utilized to evaluate model precision. It was observed that the GBDT model markedly surpassed the MLP in predictive capability and marginally exceeded both the RF and XGBoost models. Specifically, the GBDT model achieved an R$^2$ of 0.93 on the training subset and an R$^2$ of 0.88 on the validation subset. 

\setlength{\parskip}{4pt}
Prior research endeavors have similarly employed ML techniques for polymer TC prediction, utilizing a dataset of 469 polymers and decoding SMILES to a 300-dimensional feature space, yielding a prediction accuracy of R$^2$=0.828. Our model demonstrates a 5.2\% improvement in prediction accuracy over preceding research, as gauged by the R$^2$ metric on a validation set, despite utilizing a dataset of equivalent scale and a feature space reduced by a factor of 30. This achievement aligns with our projected expectations. These findings underscore the viability and promise of leveraging ML methodologies for predicting polymer TC, facilitating the identification of polymers with specific thermal conductivities, and even the discovery and creation of materials tailored for particular thermal applications.

\begin{figure}[htbp]
	\centering
	\includegraphics[width=0.5\textwidth]{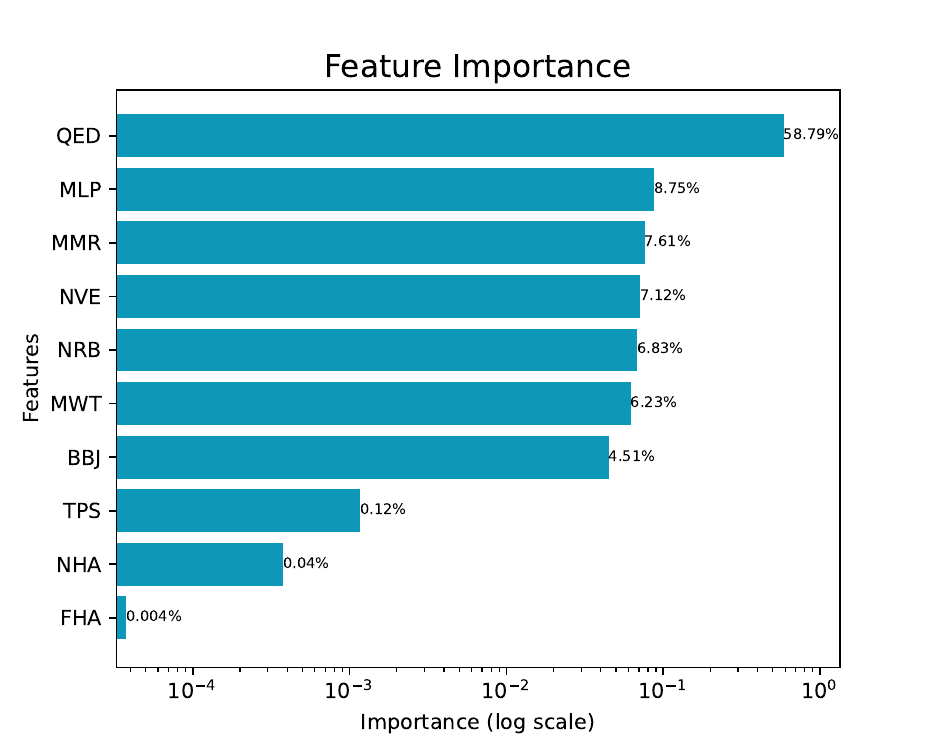}
	\caption{Bar chart of model feature importance contributions on a logarithmic scale. This figure displays the importance of SHAP features in predicting TC, as determined by the trained GBDT through SHAP feature importance analysis.}
	\label{Fig5}
\end{figure}

\setlength{\parskip}{4pt}
Each feature extracted in our study is imbued with distinct physical significance, i.e., rendering our model inherently interpretable. To delineate the impact of these features on TC and identify those of paramount importance, we employed an interpretable ML framework to generate a representation of feature importance (refer to Figure \ref{Fig5}). Employing Lundberg and Lee's SHapley Additive ExPlanations (SHAP)\cite{20}, a methodology designed to furnish interpretations for individual predictions, allows us to leverage the game-theoretical foundation of Shapley values. Here, features exhibiting substantial absolute Shapley values are deemed crucial. Aiming for a global perspective on feature importance, the figure's abscissa represents the mean of absolute Shapley values across features, while the ordinate lists the top 10 features, arranged in descending order of their SHAP importance. Within the context of the GBDT model, the QED emerges as the most influential feature, altering the average predicted absolute probability of TC by 0.5879 (0.5879 on the x-axis). Subsequently, the most significant features include MLP, MMR, NVE, and NRB, which have feature importance of 0.0875, 0.0761, 0.0712 and 0.0683, respectively. This further elucidates the validity of our method to retain features with distinct physical interpretations, such as NVE, despite their high correlations, in the preliminary phase of data analysis.

 \begin{figure}[htbp]
	\centering
	\includegraphics[width=0.5\textwidth]{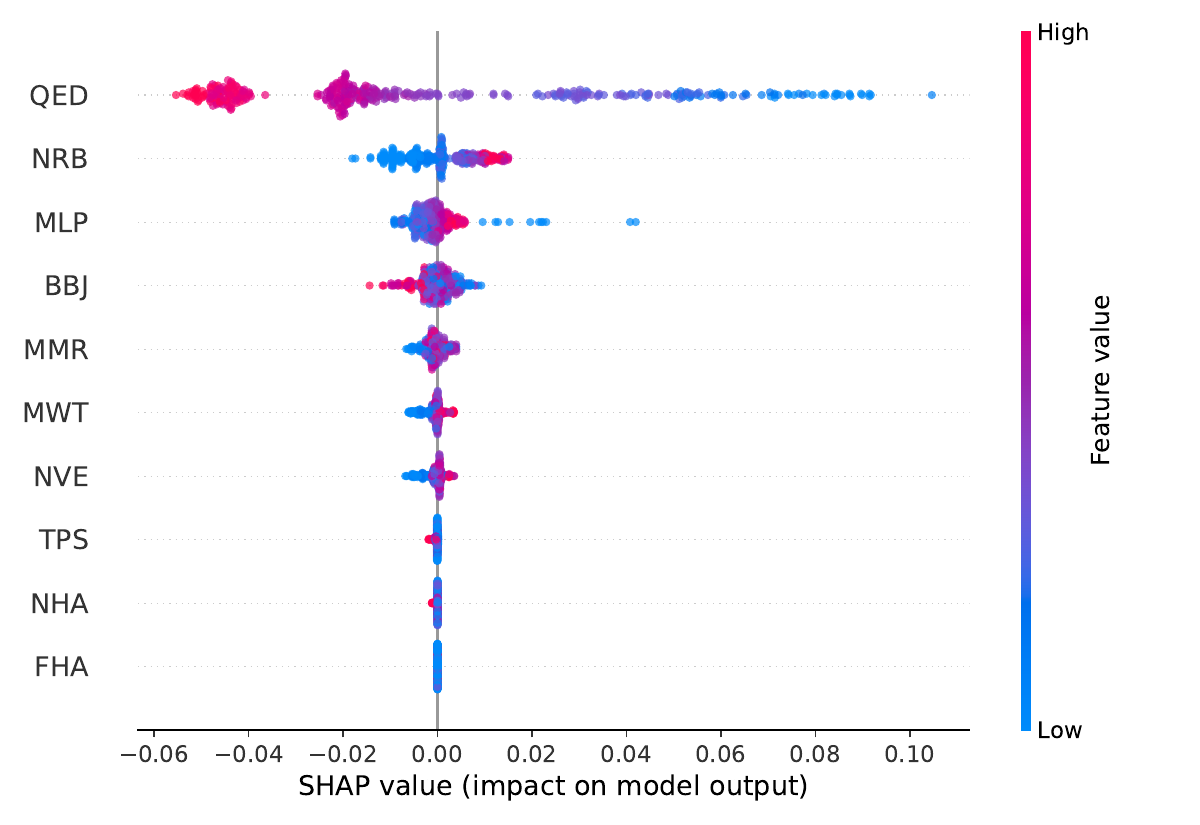}
	\caption{Average impact of model features on predictive outcomes: a SHAP summary visualization.}
	\label{Fig6}
\end{figure}

\setlength{\parskip}{4pt}
Upon determining the importance of various features, our investigation expanded to elucidate how these features—particularly the paramount ones—affect the TC. Our work aimed to decipher the physical underpinnings distinguishing the thermal conductivities of polymers characterized by disparate significant features. The SHAP summary plot (Figure \ref{Fig6}) excellently encapsulates this objective by integrating feature importance with their effects. Each dot within the summary plot corresponds to a polymer sample associated with a specific feature. The y-axis enumerates the ten distinct features, whereas the x-axis quantifies the Shapley value attributable to a feature for a given sample, with the color gradient from red to blue denoting high to low values of the feature, respectively. Within the ambit of a single feature, dots sharing identical Shapley values converge along the x-axis, and such congruent points exhibit a vertical jitter towards the y-axis. This mechanism facilitates an understanding of the Shapley value distribution for each feature.
 
\setlength{\parskip}{4pt}
Our investigation delves into the influence of key features identified within our research on the TC of polymers, aiming to unearth the physical rationales underpinning these observations. Initially, the manifestation of a positive Shapley value corresponding to a sample exhibiting a low QED value suggests that diminutive values of this particular feature positively impact the model's output. Consequently, polymers characterized by elevated QED values are inclined towards lower TC, whereas those with reduced QED values tend to demonstrate enhanced TC. To our knowledge, this constitutes the inaugural correlation of a polymer's TC with its QED value, unveiling a potential inverse relationship between the two parameters.

\setlength{\parskip}{4pt}
Polymers characterized by elevated QED values frequently exhibit complex molecular architectures, which might encompass multiple ring structures, appendant chains, or functional groups, rendering these polymers molecularly akin to pharmaceutical entities.\cite{21} It is postulated that such intricacy and the extent of branching could attenuate intermolecular forces, thereby diminishing the material's thermal energy conduction efficacy. Furthermore, an observed trend indicates that polymers with higher NRB values manifest enhanced TC, suggesting a positive correlation between NRB values and TC. This phenomenon is attributed to the premise that thermal energy transmission in polymers is contingent not solely on intermolecular interactions but also on the molecules' translational, vibrational, and rotational degrees of freedom. A substantial NRB may denote increased intramolecular degrees of freedom for absorbing and redistributing thermal energy, alongside augmented molecular flexibility to foster a more ordered structure. Under certain conditions, these molecules possessing greater specific heat capacity and inherent energy have the potential to amplify thermal transport at the macroscopic level.

\section{CONCLUSIONS}
 \renewcommand{\theequation}{5.\arabic{equation}} \setcounter{equation}{0}
In summary, we have devised a model employing ML techniques that adeptly forecasts the TC of polymers characterized by known SMILES notations. This model leverages data pertaining to the physical features and TCs of 400 polymers\cite{1, 2}. For the first time, our methodology eschews traditional text-processing tactics in favor of interpreting the SMILES notations of polymers into ten physically significant features, thereby circumventing the generation of a high-dimensional, sparse vector. The model predicts the TC of polymers on the test set with an accuracy of R$^2$=0.88. Furthermore, through the lens of interpretable analysis, we have unearthed potential inverse relationships between the TC of polymers and their QED, alongside direct correlations with NRB. These correlations are elucidated from a physical standpoint, examining factors such as intermolecular forces and molecular freedom degrees. Our model excels in identifying the traits that predicate a polymer’s TC by analyzing its monomeric units, thereby serving as a good pre-screening method in the quest for polymers of specified TCs on a grand scale. In addition, our work may provide some ideas for the design of polymers with specific TC in terms of physical properties. This ML method we designed to study the TC of polymers can also be applied in the study of other properties of polymers, which is highly generalizable and applicable.

\appendix

\end{document}